\documentstyle[twocolumn,aps]{revtex}

\begin{document}
\draft
\title{Noise in Al single electron transistors of stacked design}

\author{V.~A.~Krupenin, D.~E.~Presnov and M.~N.~Savvateev\\}
\address{Laboratory of Cryoelectronics, Moscow State University, 119899
Moscow, Russia\\}
\author{H.~Scherer, A.~B.~Zorin and  J.~Niemeyer\\}
\address{Physikalisch-Technische Bundesanstalt, D-38116 Braunschweig,
Germany\\}

\maketitle

\begin{abstract}
We have fabricated and examined several Al single electron transistors whose 
small islands were positioned on top of a counter electrode and hence did not 
come into contact with a dielectric substrate. The equivalent charge 
noise figure of all transistors turned out to be surprisingly low,
$(2.5 - 7)\times 10^{-5}$ $e/\sqrt{\mbox{Hz}}$ at $f = 10$ Hz.  
Although the lowest detected noise originates mostly from fluctuations 
of background charge, the noise contribution of the tunnel junction 
conductances was, on occasion, found to be dominant.\\
\end{abstract}


\narrowtext


\section{Introduction}

The physics and applications of the small metallic circuits 
employing the Single Electron Tunneling (SET) have been intensively studied
in the past decade. \cite{Aver,Grab} During this study, it has become apparent 
that the fluctuations of the polarization charge
offset of small islands of an SET structure 
present a serious problem.

	It is generally agreed that such background charge noise is 
resulted from stochastic occupation of charge traps in dielectric materials 
surrounding the island and it is
usually characterized by a $1/f$ power spectrum with 
a  cut-off frequency of $f_0=$ 100$-$1000 Hz and
a magnitude of 
${{(S_Q)}^{1 \over 2} =  \delta Q_{0x} / {(\Delta f)}^{1 \over 2}} 
\sim 10^{-4}-10^{-3} \, e/\sqrt{\mbox{Hz}}$  at 10 Hz 
(see, for example, \cite{Zim,Verb,Zor1} and 
references therein). Its lowest value which has been measured by
Visscher \it {et al.} \rm \cite{Viss} is accounted for 
$7\times10^{-5} \, e/\sqrt{\mbox{Hz}}$  at 10 Hz.
At $f<f_0$ the background charge noise is dominant over the 
intrinsic SET noise and it  
dramatically reduces the performance of various devices:
electrometers 
\cite{Kor}, pumps  \cite{Kel}, traps  \cite{Krup}, etc.
For that reason, the better understanding of the nature 
of this noise and the search for ways of reducing it are very 
important for the practical realization of reliable SET devices.

	Numerous measurements of noise in SET transistors 
(i.e. the two-junction structures supplied with a gate, capacitively
coupled to its island) 
have revealed a trend towards a noise increase 
with the size of an island and, hence the contacting area 
with a "noisy" substrate. \cite{Verb,Bouch} 
Recent measurements using two SET transistors (electrometers), closely 
positioned on the same substrate, showed that their output noise signals 
were partially correlated. \cite{Zor1} This, as well as other measurements
of double \cite{Mann} and single \cite{Zimm} electrometers, again 
indicate to a possibly substantial noise contribution of the substrate. 

	The aim of this work was to eliminate this "substrate" component 
of the noise by means of sophisticated design, namely by placing a small 
metallic island of an SET transistor almost entirely on the oxidized 
base electrode.
In this configuration, the base electrode, due to its larger dimensions, 
efficiently screens the island from polarization caused by the charges
both inside and on the surface of substrate. 
Hence, the noise is mostly setted by processes in the barriers of 
tunnel junctions. Due to capability of the gate control, we are able to 
distinguish the charge fluctuations and conductance fluctuations of the 
barriers in our transistors and we show, in particular, that they can be 
uncorrelated one with the other.
Moreover, this paper reports the data which turned out to be remarkable 
for the lowest level of the background charge noise measured so far 
in SET devices. \cite{Schoe}

\section{Fabrication and characterization of the samples}

	The Al structures were fabricated on a Si substrate buffered by 
a sputtered Al$_{2}$O$_{3}$  layer 200 nm 
thick by the shadow evaporation \cite{Niem} at three different angles
(see Fig.~\ref{layout}). 
There were three  successive 
deposition cycles $in$-$situ$. After the first and 
second depositions, the Al films were oxidised and the tunnel 
barriers were thereby formed.
Particular attention was given to precise alignment of the
edges of a small Al island and a base electrode (shown by 
the thin line in the mask layout in Fig.~\ref{layout}(a)) with 
the aim of not allowing 
straightforward electron tunneling between the outer electrodes.
The top electrode, a small finger turned through 90 degrees, 
did not overlap the island 
completely, so the area of the upmost junction 
(about 60 $\times$ 40 nm$^{2}$) 
was noticably smaller than that 
of the bottom one (nominally 80 $\times$ 100 nm$^{2}$). 
By virtue of this arrangement (the top electrode overlaps the island
only partially), the electric field induced by the side 
coplanar gate electrode,  penetrated to the island and held the function of 
electrostatic control. 

	The characteristics of the sample were measured 
in a dilution refrigerator at the bath 
temperature $T$ = 30 mK. The magnetic field $B=1$ T was applied to 
suppress the superconductivity in the Al films. We used the voltage bias 
configurations in which the current $I$ was measured by op amp functioning 
as an amperemeter. This setup was characterized by an extended bandwidth 
(up to 1 kHz) \cite{Star} and a low noise floor of the order 
of 20 fA/$\sqrt{\mbox{Hz}}$ at 10 Hz.

	We have measured four samples and found their electric parameters 
(see the equivalent circuit diagram in 
Fig.~\ref{circuit-IvsVg}(a)) to be typical for metallic SET transistors. 
The total tunnel resistances 
$R_\Sigma = R_1+R_2$ were in the range 200-450 k$\Omega$, the total 
capacitances of the transistor islands $C_\Sigma = C_1+C_2+C_{\rm g}$ ranged 
from 450 aF down to 350 aF.
The gate capacitances $C_{\rm g}$ were found to be rather small 
and accounted for 0.1-0.2 aF. Due to the design features, the transistor 
characteristics were asymmetric, $C_1/C_2  \approx 3-4$ and 
$R_2/R_1  \approx 3-5$ (see Fig.~\ref{circuit-IvsVg}(b)). Here 
index 1 is associated with the bottom, i.e. 
the larger, junction and index 2 with the top junction. 
Due to rather small $C_\Sigma $, the maximum values of the
current-to-charge ratio 
$ \eta = \max{_{Q_0}}{{\left|{\partial I 
    \over\partial Q_0}\right|}_{V=const}}, $ where 
$Q_0 = C_{\rm {g}} V_{\rm {gate}}$,  
were sufficiently large, especially for the steeper slopes of the 
$I-V_ {\rm {gate}}$ curves:  $\eta_{\rm {steep}} \approx 2-3 $ nA/$e$, and this 
has substantially improved the signal-to-noise ratio of our 
transistors as electrometers.   

\section{The results of noise measurements}

	The equivalent charge noise ${(S_Q)}^{1 \over 2} = 
{\eta}^{-1} {(S_I)}^{1 \over 2}$ 
(here $S_I$ is the output noise power) measured 
in all four samples at low currents turned
out to be suprisingly low (see Table \ref{table1}).  
In particular, the best charge sensitivity of the electrometer (sample 1) 
at 10 Hz was found to be 
$(2.5 \pm 0.5)\times10^{-5}$ $ e/\sqrt{\mbox{Hz}}$ or, in energy units,  

$$ \epsilon = 
{(\delta Q_{0x})^2 \over 2C_{\Sigma}\Delta f} = 
{S_I \over 2C_{\Sigma}}
\approx 230\ \hbar.	\eqno (1) $$
\\Note, that this level is still considerably higher than the fundamental noise 
floor, which we evaluate \cite{Kor} (assuming the effective 
electron temperature of the island of 100 mK) as 
$\sim 3\times10^{-6}$ $ e/\sqrt{\mbox{Hz}}$ or  $\epsilon \sim 4 \, \hbar$.
However, the obtained noise figure Eq.(1) is substantially better than 
the best one obtained earlier by Visscher \it et al. \rm \cite{Viss}: 
$7\times10^{-5}$ $ e/\sqrt{\mbox{Hz}}$ or 
$1000 \, \hbar$ at 10 Hz. Note that desing of their samples with gates
positioned beneath the islands, also assisted in partial electric screening
of the islands from the substrate charges.  

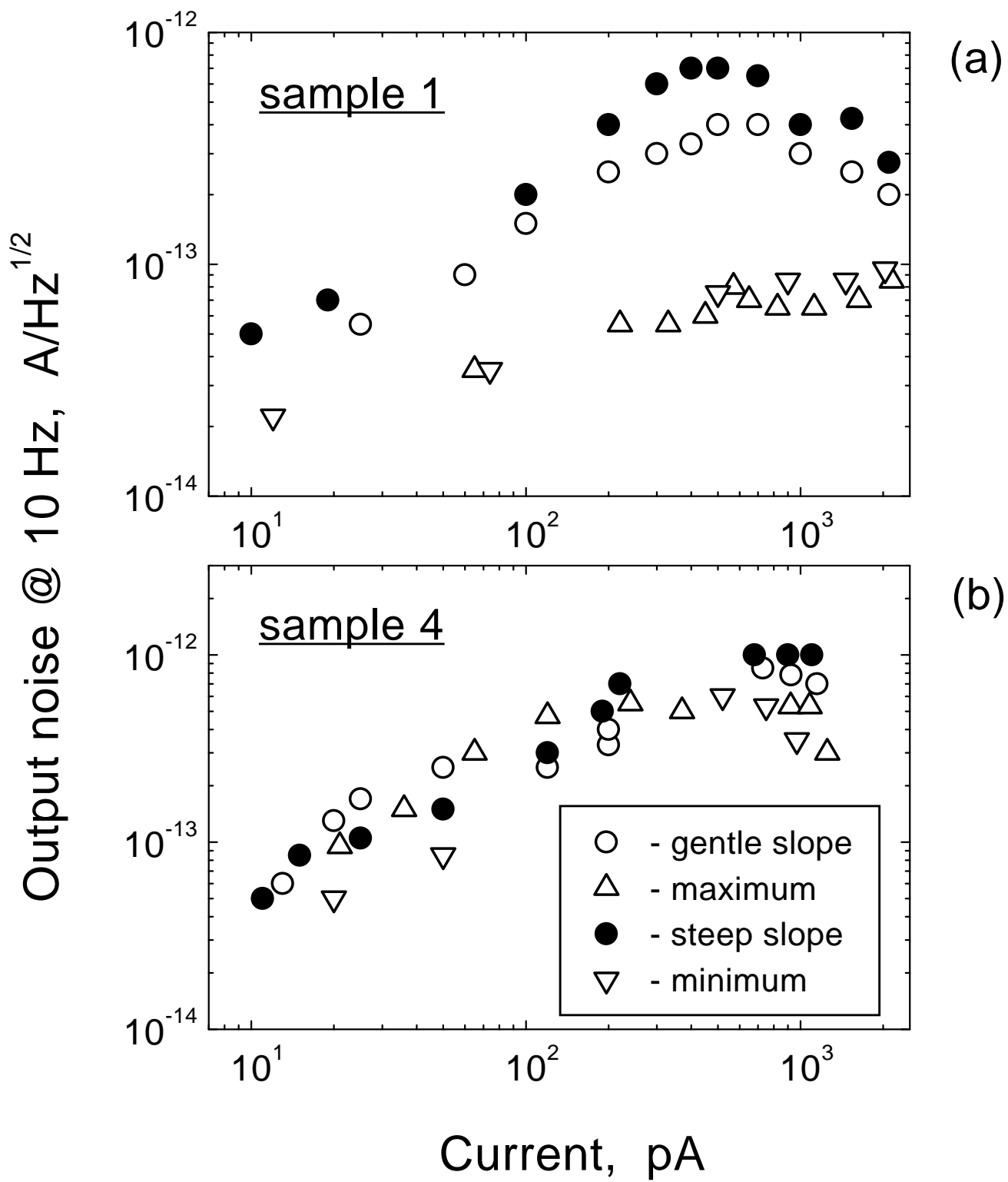
\begin{figure}
\caption{Output noise as a function of average current in different working
points of the $I-V_ {\rm {gate}}$ curve for the regular sample 1 (a) and 
the anomalous sample 4 (b).}
\label{SvsI}
\end{figure}

	The dependence of noise on the transport current (for the fixed 
voltage regime) and on the position of the bias point (marked in 
Fig.~\ref{circuit-IvsVg}(b)) is presented in Fig.~\ref{SvsI}. 
Panel (a) shows a regular dependence on the bias point by the example of 
sample 1 (samples 2 and 3 showed a similar behavior). 
One can see that output noise, measured on the slopes of the modulation 
curve appreciably exceeds the noise measured in the points of minima and 
maxima, which are insensitive to the background charge noise, $\eta=0$. 
The noise power on the slopes first rises with current, although 
starting from approximately $I= 500$ pA it 
reduces. Such behavior at high currents is attributed to the 
reduction of the current-to-charge ratio $\eta$. The equivalent charge 
noise measured, for example, on the steep slope therewith rises 
monotonously from $2.5\times10^{-5}$ $ e/\sqrt{\mbox{Hz}}$
at 10 pA up to $6\times10^{-4}$ $ e/\sqrt{\mbox{Hz}}$ 
at 2 nA.

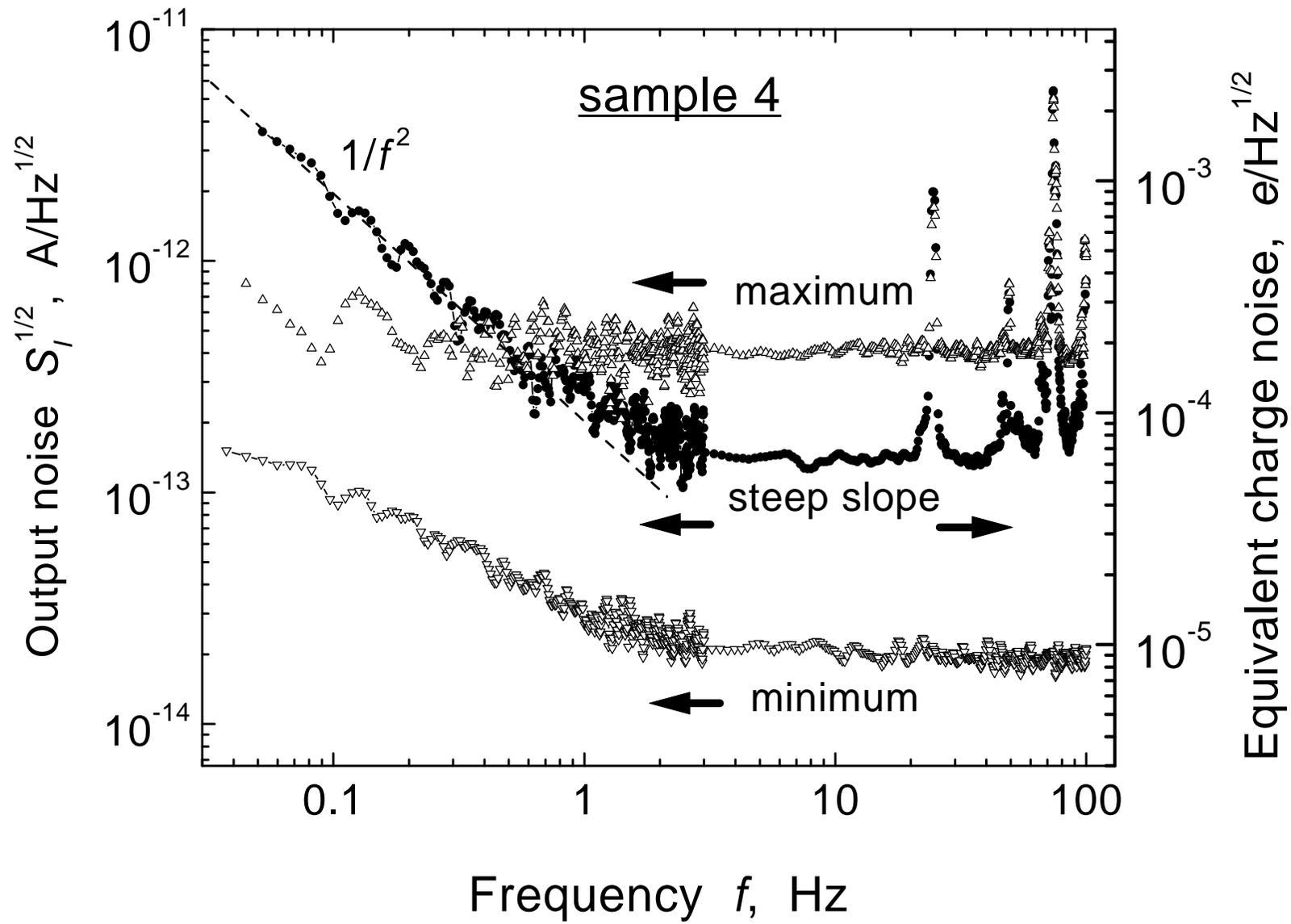
\begin{figure}
\caption{Spectrum of the output noise measured in sample 4 at different 
bias points. The noise measured at $f > 1$ Hz in the point, most 
sensitive to the background charge fluctuations (steep slope), is below
the noise in the "charge-insensitive" point (maximum). 
Note that the shot noise level, ${(S_I)}^{1 \over 2} = 
{(2eI)}^{1 \over 2}$, is below 
$10^{-14} $\, A/Hz$^{1 \over 2}$ in every bias points 
and, hence, this noise does not account for the observed spectra.}
\label{spectr}
\end{figure}

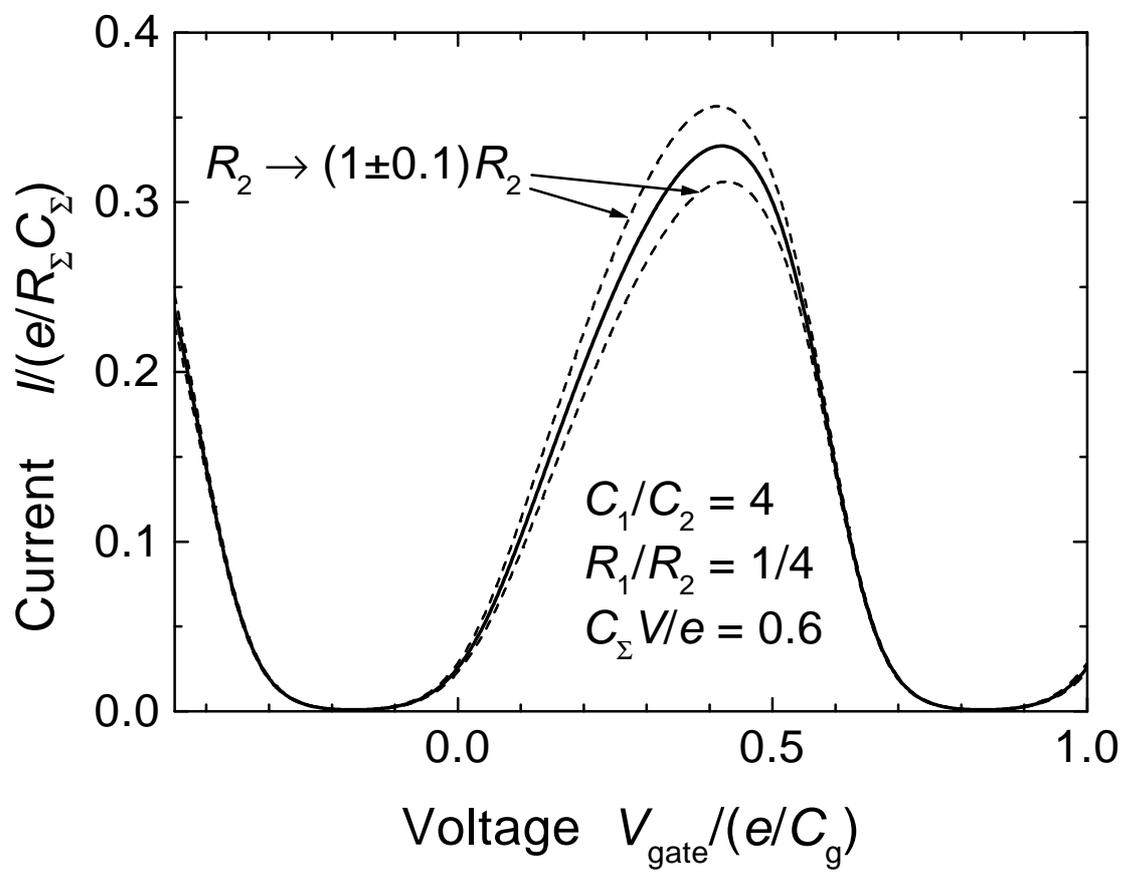
\begin{figure}
\caption{The simulated $I-V_{\rm {gate}}$ curves showing
the effect of the $\pm $10\% variation of the larger resistance $R_2$ 
(corresponding to the upper junction). Since junction 2 has smaller 
dimensions, $R_2$ is more subjected to fluctuations due to spatial
non-uniformity of the barrier, if all other 
factors being the same. One can see 
that fluctuations on the gentle slope are larger than those on the
steep slope and they rise with $I$ approaching their maximum
near the top of the curve. This agrees qualitatively with the 
observations of noise in sample 4 at $f>$1 Hz.}
\label{simul}
\end{figure}

	On the contrary, sample 4 clearly shows an anomalous 
behavior (see Fig.~\ref{SvsI}(b)) .
Its noise at $f = 10$ Hz  measured in maxima is almost similar 
to that on the slopes in a wide range of $I$, except high currents, 
$I \approx 1$ nA. 
This is in contrast to other our samples and to the observations 
in conventional SET transistors (see, for instance Refs. \cite{Verb,Star}). 
The anomalous behavoir of sample 4
is also traced in Fig.~\ref{spectr}, which demonstrates the 
power spectra measured at 
fixed voltage $V = 50 \> \mu$V, but at different transport currents. 
As can be seen, the noise
on the steep slope at $I = 50$ pA has approximately a $1/f^2$ spectrum 
for $f$ below several Hz and it dominates 
over the noise measured in the maximum ($I=100$ pA) 
at $f<1$ Hz. At higher frequencies, $f > 1$ Hz, noise 
in the maximum is dominant. 
Its spectrum is almost flat in the frequency range 0.1-100 Hz.
(Unfortunately, considerable noise of the setup at higher 
frequencies did not allow us to investigate its rolloff.)
Note, that noise, measured on the gentle slope at the same current 
$I=50$ pA (not shown), slightly exceeds the noise on the steep slope, 
but is also lower than the noise in the maximum at $f<1$ Hz. 
These observations clearly indicate two different sources of noise. 
The low-frequency component of 
noise on the slopes can be unambiguously associated with the background 
charge fluctuations, whereas noise in maxima, as well as on both slopes 
at $f > 1$ Hz, might be attributed to the tunnel conductance fluctuations
which resulted in an output noise rising with $I$ (see Fig.~\ref{simul}).

\section{Discussion}

	Thus, our measurements clearly demonstrate coexistance of two 
types of noise in the double tunnel junctions: the background charge noise 
and the conductance noise, and these two noises are characterized 
by different frequency spectra and, hence seem to be uncorrelated.  
The former is associated with 
activity of fluctuators with characteristic switching time of the order of a 
second or longer. The latter is produced by a faster switching of 
conductance of the barriers. Earlier, a possible mechanism of the 
conductance noise 
observed in single Nb-Nb$_{2}$O$_{5}$-PbBi tunnel junctions of small 
size was proposed by Rogers and Buhrman. \cite{Rog} They explained this 
(usually $1/f$) noise by stochastic charging and discharging of, at least, 
several natural traps (two-level fluctuators) for single electrons, which are 
located in the tunnel barriers. According to their picture, if  an electron 
is captured by the trap, it repels other electrons attempting to tunnel nearby.
Such trap causes barrier height fluctuations and, hence, it should 
unavoidably produce a polarization of the electrodes which form this 
junction. However, their samples (single junctions) were evidently 
insensitive to this polarization. 
In our case of a double junction a (noticable) polarization of  
a small island could then be apparently developed. Our 
observations, however, do not 
reveal this effect and hence they cast doubt on the mechanism based on 
re-charging of traps, located inside the barrier, 
as the main mechanism responsible for the conductance fluctuations
in our Al-Al$_{2}$O$_{3}$-Al junctions. What could be an 
alternative mechanism of "pure" conductance noise?
As a possible mechanism, we suggest that the observed 
conductance noise originates from fluctuators which are located 
immediately in the metal-oxide interfaces and influence the local current 
density but do not polarize the island. 
In our case of a thin island ($d=$15 nm) the inner boundaries of the island 
experience multiple reflections of an electron arrived to the island before 
its thermalization. (This process is characterized by the inelestic scattering 
length $\sim \mu$m $\gg d$.) 
In the course of these reflections the confined electron can donate its energy 
to those ions on the metal-oxide boundary which are not constrained to 
certain positions in the lattice, causing thereby their slow motion between 
neighboring minima. This process can possibly cause fluctuations in the 
effective barrier transparancy and hence in the tunnel current.

	However, for the most part of the measured samples
the conductance noise did not manifest itself. 
(This agrees with the fact that for the perfect 
Al-Al$_{2}$O$_{3}$-Al barriers the measured noise is 
usually close to shot noise level. \cite{Spea}) 
In these cases, we could assume that the 
observed background charge noise and, possibly, very rare 
switching of $Q_0$ originate from re-charging of traps, located 
inside the barrier. On the other hand, a motion of charges in the natural 
oxide layer covering the open surface of the island and  having more 
irregular structure can also be a source of background charge noise. 
These areas do not directly contribute to electron current through 
the device and hence do not cause conductance fluctuations. Moreover, 
since the observed perfect Coulomb blockade did not show any sign of a 
parallel channel for electrons to tunnel between the outer electrodes, we 
do not rule out the existence of small areas where the edge of an island is 
in contact with the substrate, that can also contribute to the total noise. 
This is in confirmity with our observation of substantially larger charge 
noise in the reference transistors which were fabricated on the same 
chips such that their islands were partly lying on the substrate. For example, 
two transistors from the same chip as sample 2 had a contacting area with 
the substrate of about 20\% and 50\% of the total island area and they 
exhibited a noise of  $1.2 \times 10^{-4} \, e/\sqrt{\mbox{Hz}}$ and 
$2 \times 10^{-4} \, e/\sqrt{\mbox{Hz}}$ at $f=10$ Hz 
respectively \cite{Krup1}
(compare with the bare value of  $4 \times 10^{-5} \, e/\sqrt{\mbox{Hz}}$ 
presented in Table \ref{table1}). 

	Finally, we have shown that, first, the achieved low level of noise in 
our transistors is definitely due to their stacked design which eliminates the 
effect of a substrate. Secondly, the 
noise of the barrier conductances which was observed in one of such
samples, was clearly resolved since it dominated over the background 
charge noise at $f >$ 1 Hz.
Although the mechanism of this conductance noise is not completely 
understood it might be associated with fluctuations in the metal-oxide
interface which almost do not cause polarization of the island. 
As long as this noise is much lower than the typical background charge
noise level,
it seems not to be a hindrance for majority of SET devices, 
operating at low current. 
Therefore, our experiment encourages us to further study of noise both in 
devices of traditional design fabricated with new materials for the substrate 
as well as in devices having an alternative (e.g., stacked) design with a 
goal to realize the high potentials of metallic SET structures.

\acknowledgments

	We gratefully acknowledge the contributions from 
A.~B.~Pavolotsky, S.~V.~Lotkhov and U.~Becker and the stimulating 
discussions with B.~Starmark on the setup electronics. The work is 
supported in part by the Russian Scientific Program "Physics of Solid State 
Nanostructures", the Russian Fund for Fundamental Research, the German 
BMBF and the EU (SMT Research Project SETamp).

\begin{figure}
\caption{The schematics of the mask (a), which assumes precise 
alignment of the edges of an island and a base electrode (shown 
by thin line). An SEM image (b) of the structure resulted
from successive three angle evaporation of Al and two
oxidations (after the first and second depositions).}
\label{layout}
\end{figure}

\begin{figure}
\caption{(a) The circuit diagram of the SET transistor where small
tunnel junctions are marked by the devided rectangular boxes.
(b) The $I-V_ {\rm {gate}}$ curve (sample 1, $V = 300 \> \mu$V) where 
positions of the working points for the noise measurements are marked.}
\label{circuit-IvsVg}
\end{figure}

\begin{table}
\caption{Equivalent charge noise 
in four stacked SET transistors at $f=10$ Hz 
measured at low currents ($I \approx 10-20$ pA).} 
\label{table1}
\begin{tabular}{ccccc}
Sample \#&1&2&3&4\\
\tableline
${(S_Q)}^{1 \over 2} \ [{e \over \sqrt{\mbox{Hz}}}]$&
$2.5 \times 10^{-5}$&$4 \times 10^{-5}$&
$7 \times 10^{-5}$&$5 \times 10^{-5}$\\
\end{tabular}
\end{table}

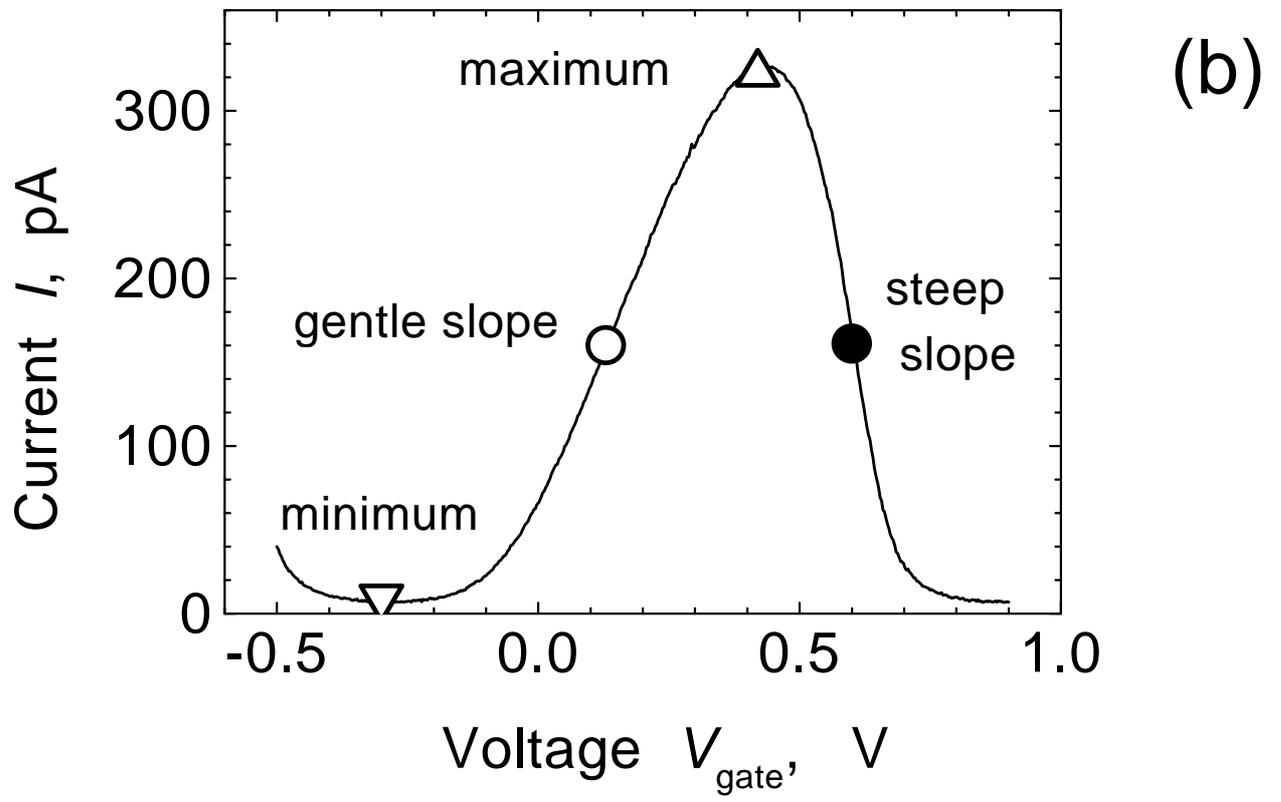
\end{document}